\documentclass[twocolumn, prl, showpacs, aps]{revtex4} 
\usepackage{graphicx}
\usepackage{amsmath}
\usepackage{amsbsy}
\usepackage{mathbbol}

\graphicspath{{figures/}}

\begin{document}

\title{Mesoscopic fluctuations in the spin-electric susceptibility due to Rashba spin-orbit interaction}

\author{Mathias Duckheim}
\author{Daniel Loss}
\affiliation{Department of Physics University of Basel, CH-4056
Basel, Switzerland} 

\date{\today}
%





\begin{abstract}
  We investigate mesoscopic fluctuations in the spin polarization generated by a static electric field
  and by Rashba spin-orbit interaction in a 
  disordered 2D electron gas.  In a
  diagrammatic approach we find that the out-of-plane polarization --
  while being zero for self-averaging systems -- exhibits large
  sample-to-sample fluctuations which are shown to be well within
  experimental reach.  We evaluate the disorder-averaged variance of
  the susceptibility and find its dependence on magnetic field,
  spin-orbit interaction, dephasing, and chemical potential
  difference. \end{abstract}

\maketitle
A primary goal in semiconductor spin physics is the control of
magnetic moments by electric fields \cite{spintronics, awschalom:07}.
One way to achieve this is to make use of the magnetoelectric effect
(MEE) \cite{Levitov1985,Edelstein1990}, a spin polarized steady state
which emerges from intrinsic 'magnetic' fields generated by spin-orbit
interactions (SOI) and transport.  While this MEE has been observed
e.g. in n-InGaAs epilayers \cite{Kato2004a,Kato2004,Kato2005} and hole
gases \cite{Silov2004}, the resulting net polarization is below
percent for electron systems \cite{Kato2004a,Kato2005}, and thus much
smaller than what has been achieved by optical pumping
\cite{Kato2004,Stich2007,Meier2007}.  Moreover, in a standard
two-dimensional electron gas (2DEG), with typical Rashba SOI
\cite{Bychkov1984}, the MEE generates only in-plane spin polarization
but no out-of-plane components
\cite{Edelstein1990}. The latter would be desirable, also since they
can be detected more easily e.g. by optical means.

However, these observations apply only to disordered phase-incoherent
systems with self-averaging \cite{Akkermans2007}. On the other hand,
it is well-known that phase-coherence in mesoscopic systems leads to
new quantum effects such as conductance fluctuations or weak
antilocalization, especially due to intrinsic SOI \cite{Altshuler1985,
  Altshuler1991a, Lee1987, Zumbuehl2005,Aleiner2001,Miller2003}.
 
Similarly, mesoscopic spin effects emerge for the MEE when the system
becomes phase-coherent.  Indeed, focussing on 2DEGs with Rashba SOI,
we will show here that the spin-electric susceptibility is subject to
strong sample-to-sample fluctuations, and thus individual mesoscopic
samples with phase coherence can exhibit a large net spin
polarization. Quite remarkably, these strong fluctuations show up not
only in the in-plane but also in the out-of-plane spin components.  We
find that these fluctuations considerably exceed the polarization
obtained in self-averaging samples, and since the latter has been
successfully measured by optical means \cite{Kato2004a,Kato2005}, the
spin fluctuations predicted here should be well within experimental
reach.  We will see that this strong enhancement is special for Rashba
(or Dresselhaus\footnote{We note that Dresselhaus and Rashba SOI give
  the same results since the corresponding Hamiltonians are unitarily
  related \cite{Chalaev2005}.})  SOI, and, diagrammatically, it
results from a spin vertex renormalization typical for such intrinsic
SOI \cite{Edelstein1990,Chalaev2005,Duckheim2006}.

Related effects studied before are local spin fluctuations in metallic
conductors due to the extrinsic spin-orbit effect \cite{Zyuzin1990}
(as opposed to the intrinsic Rashba SOI \cite{Engel2006}), density of
states fluctuations of quantum corrals \cite{Walls2007}, and
fluctuations of spin currents in general nanostructures
\cite{Bardarson2007} and chaotic quantum dots
\cite{Bardarson2007,Krich2008}.  While thermal spin fluctuations have
been observed \cite{Oestreich2005}, we are not aware of studies of
mesoscopic spin fluctuations due to the MEE as described here.


\begin{figure}[t]
  { \includegraphics[width = 0.4  \textwidth]{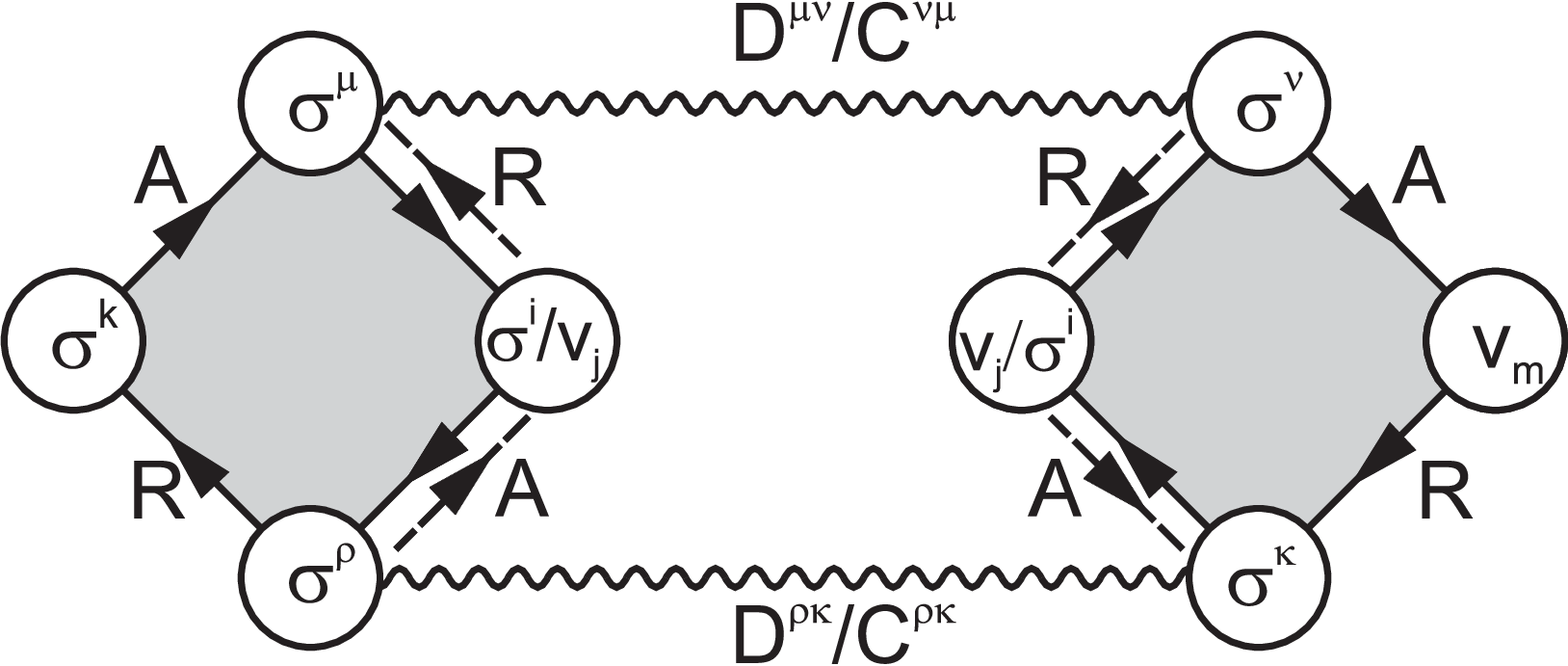}
   \label{fig:a_fluctuations-diff}    } \\
\vspace{0.2cm}
  {\includegraphics[width = 0.4  \textwidth]{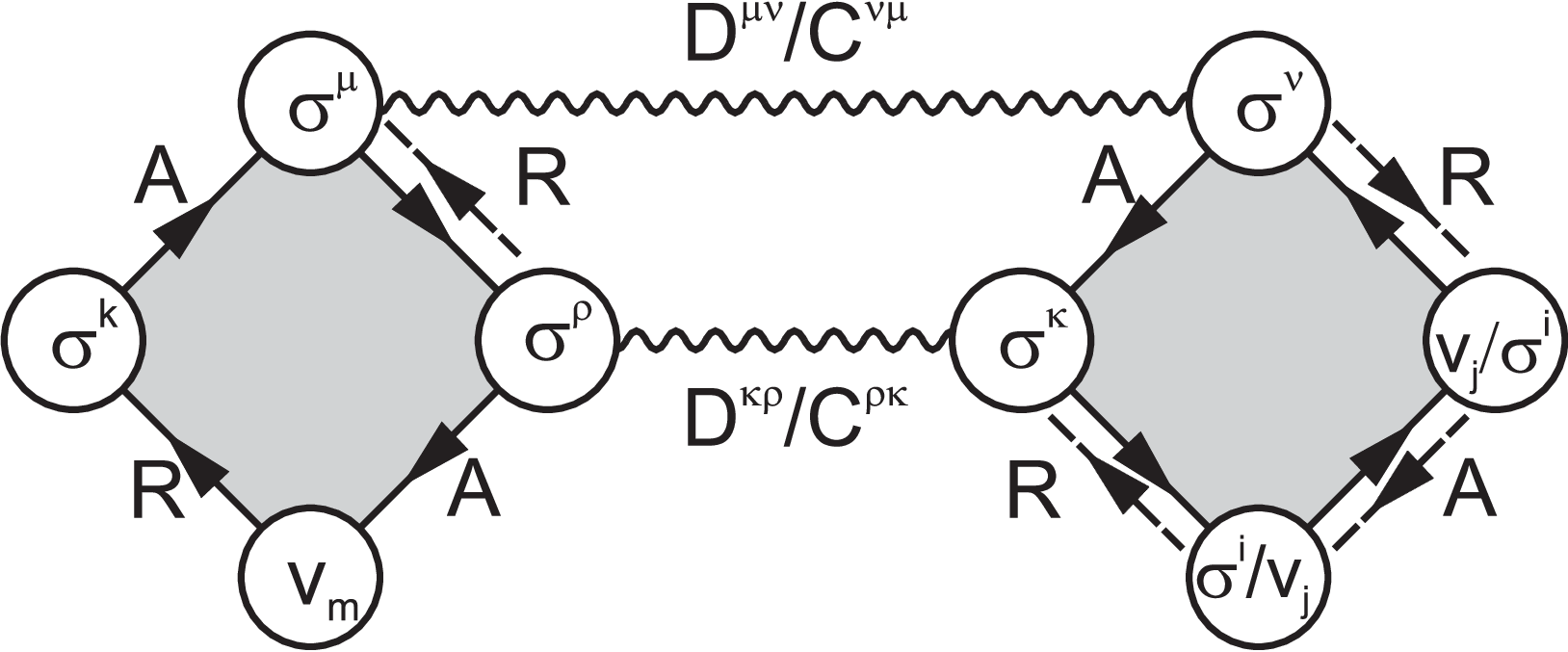}
   \label{fig:b_fluctuations-dens-states}    } \quad

 \caption{Dominant diagrams (for $1/p_{F}l\ll 1$) leading to the
   variance $\overline{\delta \chi^{ij}\delta \chi^{km}}$ given in
   Eq.~(\ref{eq:fluctuations}), see also Fig.~\ref{fig:HB-general}.
   The upper diagram contains the Hikami boxes (HB) $V_{La},
   V_{Ra}$(with solid arrows and two Diffusons (D)) and $V_{Ld},
   V_{Rd}$ (with dashed arrows and two Cooperons (C)), {\it resp.}
   The lower diagram contains the HBs $V_{Lb}, V_{Rb}$ (solid arrows,
   two Diffusons) and $V_{Lc}, V_{Rc}$ (dashed arrows, two Cooperons).
 }
 \label{fig:fluct-diagrams}
\end{figure}

We consider a disordered mesoscopic square-shaped 2DEG of size $L^{2}$
containing non-interacting electrons of mass $m$ and charge $e$ and
described by the Hamiltonian
\begin{align}
\label{eq:hamiltonian}
H = \frac{\mathbf p^2}{2 m } + \alpha (p_1 \sigma^2 - p_2 \sigma^1) + \mathbf
b \cdot \boldsymbol{\sigma} + V.
\end{align}
Here,  $\mathbf p = (p_1, p_2,0 )$ is the in-plane momentum, $\alpha$
 the Rashba SOI constant \cite{Bychkov1984}, $2b = g \mu_B B
(\cos \varphi_B,\sin \varphi_B,0)$ an external in-plane magnetic
field, and $\boldsymbol{\sigma} = (\sigma^1, \sigma^2, \sigma^3)$ the
Pauli matrices (and $\sigma^0 = \mathbb{1}$). The disorder potential $V$ is due to static short-ranged
impurities randomly distributed  and  characterized
by the mean free path $l = \tau p_F/m $, where $\tau$ is the 
scattering time and $p_F$ the Fermi momentum.

The spin polarization due the MEE is given in linear response by $
\langle \sigma^i \rangle = \chi^{ij} E_j$, $i=1,2,3$, where $E_{j}$ is
a static electric field applied along the j-direction and $\chi^{ij}$
the (zero-frequency) spin-electric susceptibility (per unit area).
Here, we focus on the mesoscopic fluctuations of $\chi^{ij}$ due to
disorder, described by the variance $\overline{ (\delta
  \chi^{ij})^2}$, where $\delta \chi^{ij} = \chi^{ij} - \overline{
  \chi^{ij}} $ and where the overbar denotes disorder averaging.  We
start from the Kubo formula for $\chi^{ij}$ expressed in terms of
retarded/advanced Green functions $G^{R/A}_{E_F}$ at the Fermi energy
$E_{F}$ \cite{Duckheim2006}
\begin{align}
  \label{eq:kubo}
  \chi^{ij}(E_F) = \frac{e}{4 \pi} \mathrm{Tr}  \sigma^i (G^R_{E_F}
  - G^A_{E_F}) v_j (G^R_{E_F} - G^A_{E_F}) \, ,
\end{align}
where $\mathrm{Tr} \rightarrow \int d^2p/(2 \pi)^2 \mathrm{tr}_S$
denotes momentum integration and spin trace, and $v_j=i [H, x_j]/\hbar
$ is the SOI-dependent velocity operator. In Eq.~(\ref{eq:kubo}) we
have used time reversal invariance \footnote{The B-field breaks
  time-reversal invariance and leads to additional terms in
  Eq.~(\ref{eq:kubo}).  However, their contribution to 
  $\overline{(\delta \chi^{ij})^2}$ is negligible for the small
  B-fields $b \tau \approx \sqrt{x^2 \phi}$ relevant for C, see
  Eq.  (\ref{eq:mag-field-supp}).}  to make the symmetry $\chi^{ij} =
\chi^{ji}$ explicit\footnote{Note that $G^{R} G^{R}$ ($G^{A} G^{A}$)
drops out in Eq.~(\ref{eq:kubo}) due to the identity $ G^{R/A}_E v_j G^{R/A}_E = -i
[x_j, G^{R/A}_E] /\hbar$.}.

The variance $\overline{(\delta \chi^{ij})^2}$ is obtained as the
impurity average over the product of two susceptibilities given in
Eq.~(\ref{eq:kubo}). Extending the diagrammatic approach of \cite{Altshuler1991a,Edelstein1990,Akkermans2007}
to include SOI and spin vertices, we obtain the diagrams shown in
Fig.~\ref{fig:fluct-diagrams} which give the dominant contribution to
the variance for  $1/p{F}l\ll 1$. Explicitly, we find

\begin{align}
  \label{eq:fluctuations}
\overline{\delta \chi^{ij}} &\overline{(E_F + \Delta ) \delta \chi^{km}(E_F)}
 =   \left(\frac{e}{2 \pi L} \right)^2 
 \int \frac{d^2 q}{(2 \pi)^2}  \Big[ \notag \\  
 V_{La}^{\mu \rho} &V_{Ra}^{\kappa \nu}   D_{-q}^{\mu
  \nu}(\Delta ) D_{q}^{\rho \kappa}(-\Delta )
   +V_{Lb}^{\mu
  \rho} V_{Rb}^{\kappa \nu}  \big\{D_{q}^{\mu
  \nu}(\Delta ) D_{q}^{ \kappa \rho}(\Delta )  
  \notag \\  +  D&_{q}^{\mu
  \nu}(-\Delta ) D_{q}^{ \kappa \rho}(- \Delta )\big\}
+   V_{Lc}^{\mu
  \rho} V_{Rc}^{\kappa \nu}  \big\{ C_{q}^{\nu
  \mu}(\Delta ) C^{\rho \kappa}(q, \Delta ) \notag \\  
  + C&_{q}^{\nu
  \mu}(- \Delta ) C_{q}^{\rho \kappa}(- \Delta ) \big\} 
  +V_{Ld}^{\mu
  \rho} V_{Rd}^{\kappa \nu}   C_{q}^{\mu
  \nu}(- \Delta ) C_{q}^{\kappa \rho }(\Delta )  \Big]  .
\end{align}
Here, $\Delta $ is the difference in chemical potentials or gate
voltages of $\chi^{ij}$ and $\chi^{km}$, and $D_{q}$ and $C_{q}$ are
the Diffuson and Cooperon matrices, resp., with $4 \times 4$
components $\mu, \nu=0,1,2,3$ (see below).  The $V_{La}$'s
($V_{Ra}$'s) are Hikami boxes (HBs) shown on the left (right) in
Fig.~\ref{fig:fluct-diagrams}.  Since $\chi^{ij} = \chi^{ji}$ (see
Eq.~(\ref{eq:kubo})), each product $V_L^{\mu \rho} V_R^{\kappa \nu}$
in Eq.~(\ref{eq:fluctuations}) turns into a sum with 4 terms.  These
terms are obtained by exchanging spin and velocity vertices in the
$V's$
such as e.g. $V_{La}^{\mu \rho} V_{Ra}^{\kappa \nu}
\equiv V_{La}^{\mu \rho}[\sigma^i, \sigma^k] V_{Ra}^{\kappa \nu}[v_j,
v_m] + (\sigma^k \leftrightarrow v_m ) + (\sigma^i \leftrightarrow v_j
) + (\sigma^k \leftrightarrow v_m , \sigma^i \leftrightarrow v_j)$.
Additionally, the  vertices have to be dressed with
non-crossing impurity lines (\cite{Duckheim2006,Chalaev2005}). In
contrast to conductance fluctuations \cite{Altshuler1991a,Akkermans2007}, such vertex corrections 
are crucial here as they give the dominant dependence on the SOI (see below). 

Let us now evaluate Eq.~(\ref{eq:fluctuations}) by calculating first
$V_{L/R}$, given in Fig.~\ref{fig:fluct-diagrams}, and then $C/D$.
From now on, we restrict ourselves to the diffusive regime $l/L \ll x
= 2 \alpha p_F \tau\ll 1$, which allows us to neglect the q-dependence
in $V_{L/R}$ and to expand in $x$.  Additionally, we may neglect $b$
and $\Delta$ in $V_{L/R}$. Indeed, we first note that $V_{b\neq0} /
V_{b=0} \propto b \tau$ and $V_{\Delta \neq 0} / V_{\Delta =0} \propto
\Delta \tau $ for small $b$ and $\Delta$, and second, that the
suppression of $C/D$ with increasing $b$ and $\Delta $ sets in on a
much smaller scale $b \tau \approx \sqrt{\phi x^2}$ and $\Delta \tau
\approx \phi$ (dephasing, see below).
\begin{figure}[ht]
   \includegraphics[width = 0.47  \textwidth]{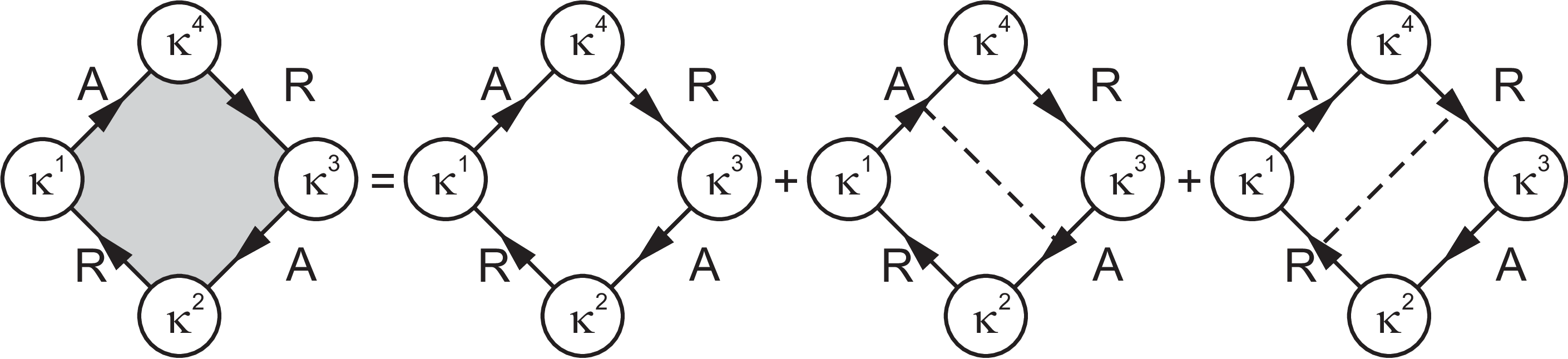}
   \caption{General Hikami box (HB): shaded area represents the correction
     to the 'empty' box $V_a$ by  $V_b, V_c$ containing a single
     impurity line (dashed). The vertices are denoted by $\kappa^i \in \{ \sigma^\mu, v_{1}, v_{2}\}$,
     and R/A stands for the averaged $G_{E_{F},b}^{R/A}$.}
\label{fig:HB-general}
\end{figure}
As a result, we can unify the calculation of the $2 \times 4 \times 4$
HBs in Fig.~\ref{fig:fluct-diagrams}. They can be expressed
by a linear relation \footnote{E.g.,  $V_{Rd}^{\kappa
    \nu}[\sigma^i, v_m] = (d k^T)^{\kappa l} (\bar d k k^T)^{i j}
  k^{\nu r} V[\sigma^l, \sigma^{j}, \sigma^{r}, v_m]$, where
  $k^{\mu l}= \mathrm{tr}\{ \sigma^2
  \sigma^\mu \sigma^l\}/2 $ and $d(\bar d) = \mathrm{diag}(1,-1,-1,-1
  ,+1(-1),+1(-1)) $.} in terms of the general HB  in
Fig.~\ref{fig:HB-general}, defined by the 'empty' box
$
\mathrm{Tr} \left\{ G^A \kappa^1 G^R \kappa^2 G^A \kappa^3 G^R
  \kappa^4 \right\}$ and the associated HBs with a single
impurity line.  Here, $\kappa^i \in \{ \sigma^\mu, v_{1}, v_{2}\}$ 
denotes vertices in Figs.~\ref{fig:fluct-diagrams}, \ref{fig:HB-general}, and 
in Eq.~(\ref{eq:fluctuations}), and  $G^{R/A}_{E_{F},b}(\mathbf p)$ is the impurity averaged Green function
which depends on B-field  and  SOI \cite{Duckheim2006}.

Next, we evaluate the  Diffuson $D$ and Cooperon $C$ 
in Eq.~(\ref{eq:fluctuations}),  given by $D_{q} = (1/2m\tau)(1 - X_{+})^{-1}$ and
$C_{q} =(1/2m\tau) k (1 - X_{-})^{-1} k$, where 
$k^{\mu l}= \mathrm{tr}\{ \sigma^2
  \sigma^\mu \sigma^l\}/2 $.
Expanding $X^{\mu \nu}_{\pm} = \mathrm{Tr} \{\sigma^\mu
G^R_{E_{F},b}(\mathbf p) \sigma^\nu G^A_{E_{F},\pm b} (\mathbf p -\mathbf q ) \}/
2 m \tau$
in $\bar{q} = q l\ll 1$, $\beta = 2 b
\tau\ll 1$, and $x$, 
we find
\begin{widetext}
\begin{align}
  \label{eq:x-insertion}
X_\epsilon = \left(
\begin{array}{llll}
\frac{1}{\lambda } -\frac{\bar{q}^2}{2} -  \beta^2 \delta_{\epsilon,-1}  & -
i \beta \delta_{\epsilon,-1}   & 0 & 0 \\
 - i \beta \delta_{\epsilon,-1}   & \frac{1}{\lambda } -\frac{\bar{q}^2}{2} -\frac{x^2}{2} -
   \beta^2 \delta_{\epsilon,-1} & 0 & i x \bar{q}  \cos \varphi  \\
 0 & 0 &\frac{1}{\lambda } -\frac{\bar{q}^2}{2} - \frac{x^2}{2}
 -\beta^2 \delta_{\epsilon,+1} & -\beta \delta_{\epsilon,+1}   + i x \bar{q}  \sin \varphi  \\
 0 & -i x \bar{q}  \cos \varphi  & \beta \delta_{\epsilon,+1} - i
 x \bar{q}  \sin \varphi  &\frac{1}{\lambda } -\frac{\bar{q}^2}{2} -x^2
 -  \beta^2
 \delta_{\epsilon,+1}
\end{array}
\right) \, ,
\end{align}
\end{widetext}
where $\varphi$ is the polar angle of the momentum $\bar{\mathbf q}$
and  $\lambda= 1 - i \Delta  \tau$. 
We will find below that the main
contribution to $\overline{\delta \chi^2}$ comes from the $1/\bar q^2$
terms in $D^{00}$ and $C^{33}$.  To account for orbital
dephasing  we introduce a corresponding dephasing time
$\tau_\phi$ in $D$ and $C$ by the standard replacement (see e.g.
Eq.(3.15) in \cite{Bergmann1984})  $\Delta  \rightarrow \Delta  +
i/\tau_\phi$.

Although generalized to include spin vertices, the method presented so
far involves the calculation of similar diagrams as for the conductance
fluctuations \cite{Altshuler1985,Akkermans2007}. An important difference, however, is the inclusion of
the vertex corrections for spin and spin-dependent velocity which we discuss
now. The vertex correction is an infinite sum of diagrams which
consists of non-crossing impurity lines connecting 
the retarded and
advanced Green functions in either $\chi^{ij}$ or $\chi^{km}$,
{\it resp.} For the velocity vertex this leads to $v_j \rightarrow
p_j/m$, i.e.  the spin part of the velocity is cancelled in
the dc limit \cite{Chalaev2005}. For the spin vertex this leads to the
replacement $\sigma^i \rightarrow \Sigma^{i\mu} \sigma^\mu$ where
$\Sigma$ is diagonal given by $\Sigma = (1 - X_+)^{-1}$ at $\bar q =0$, with the
relevant entries $\Sigma^{11} = \Sigma^{22} = 2 / x^2$ and
$\Sigma^{33} = 1/x^2$. These expressions are valid in the regime $l/L
\ll x$ and will be used here. For general $x,L$, the finite size form
of the vertex correction has to be taken into account, giving e.g.
$\Sigma^{22} = (2/x^2) (1 - \tanh{(x L/2 l)}/ (x L/2 l) )$, which then renders
Eq.~(\ref{eq:variance-no-b}) given below finite for $x \to 0$.

In the regime $\mathrm{max}\{(l/L)^2, \tau/\tau_\phi\} \ll x^2$, we find
for the in-plane ($i = 1,2$) and  out-of-plane ($i = 3$)
components  of the variance
\begin{align}
\label{eq:fluctuations2}
&\overline{(\delta \chi^{ij})^2} = \left( \frac{e\delta \bar q^2}{8 \pi^3 v_F}\right)^2
  \sum_{n_x=1, n_y=0}^\infty
 s^{ij} ( \bar q_{n_x},\bar q_{n_y}) \, ,
\end{align}
where $v_F =p_F/m$ is the Fermi velocity, $\delta \bar q = \pi l/L$,
and the sum over the $\bar q_{n} \equiv \delta \bar q n$ satisfies the
mixed boundary conditions  for the Diffuson \cite{Akkermans2007} in a finite  sample
of square size  $L$ with two opposite sides  attached to the leads. Here, 
$s^{ij}$ in Eq.~(\ref{eq:fluctuations2}) are rational functions in
$\bar q_{n_x}, \bar q_{n_y}$ depending parametrically on $x$, $\mathbf
b$, $\Delta $, and the orbital dephasing rate $\phi = \tau/\tau_{\phi} = 2
l^2/L_\phi^2$.

Evaluating $s^{ij}$ for $l/L, l/L_\phi \ll x$ first for $b=
\Delta  =0$, and choosing   the E-field
along x  (i.e. $j=1$) we find
\begin{align}
  \label{eq:limites}
  s^{i1} = \frac{a_i\left(4x\Sigma^{ii} \right)^2}{(\bar q^2 + 2 \phi)^2}\, ,
\end{align}
where $a_1 = a_2 = 1$ and $a_3 = 2$.
From Eqs.~(\ref{eq:fluctuations2}), (\ref{eq:limites}) we then obtain
$\overline{(\delta \chi^{i1})^2} = (ex\Sigma^{ii}/2 \pi^{3} v_F)^2 c_2
 a_i $, where $c_2 = \sum_{n_x,n_y} \delta
\bar{q}^4/[(n_x^2 + n_y^2) \delta\bar{q}^2 + 2 \phi ]^2$. 
To assess the
magnitude of this result we compare it to the average of the in-plane
susceptibility $\overline{\chi^{21}} = ex/2 \pi v_{F}$  \cite{Edelstein1990}. For
the out-of-plane component of the spin fluctuations this yields
\begin{align}
  \label{eq:variance-no-b}
  \frac{\overline{(\delta \chi^{31})^2}}{(\overline{\chi^{21}})^2} = \frac{2
    c_2}{\pi^4 x^4} = \frac{c_2}{2 \pi^4} \frac{\tau_{DP}^2}{\tau^2} \, ,
\end{align}
and similarly for the in-plane components. Thus, we see that the
relative fluctuations grow with  increasing $\tau_{DP}$, where $\tau_{DP} = 2 \tau / x^2$
is the D'yakonov-Perel spin
relaxation time 
\cite{Dyakonov1972}.


  Remarkably,  for
negligible dephasing, i.e. $\phi \ll 1$, one obtains $c_2 \approx
1.51$, and the spin fluctuations $\overline{(\delta \chi^{ij})^2}$
become independent of the sample size. Typical numbers for a GaAs 2DEG
(that are consistent with the regime of validity for $L_\phi = 10 l $)
yield $x=0.1 \dots 1$ and thus result in large relative  fluctuations, 
$\sqrt{\overline{(\delta
    \chi^{ij})^2}} / \overline{\chi^{21}} \approx 20 \dots 0.2$.
In other words,
there exist specific impurity configurations and realistic system parameters that give
rise to a large out-of-plane spin polarization in
response to a static in-plane electric field.
\begin{figure}[hb]
  \includegraphics[width = 0.45 \textwidth]{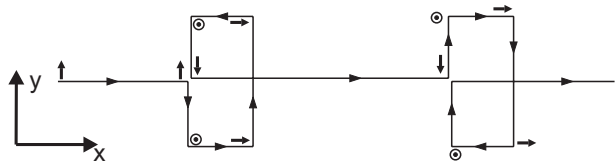}
  \caption{Example of a propagation path along which  an initially
    y-polarized spin (small arrow)  rotates exclusively  into the positive
    out-of-plane z-direction (denoted by circled dots). The spin
    directions are denoted by small arrows at the left of each segment
    and result from precession about the instantaneous Rashba
    spin-orbit field. }
\label{fig:extremal-path}
\end{figure}

To gain physical insight into this result, we consider an electron
with spin initially pointing along the $y$-axis, see
Fig.~\ref{fig:extremal-path}.  While the electron propagates
coherently through the sample, the spin precesses about the intrinsic
Rashba SOI field which is in-plane and perpendicular to the
propagation direction. As a result of orbital phase coherence
the electron propagates along a
path that is preferred by constructive interference in the given
disorder configuration.  Fig.~\ref{fig:extremal-path} shows an
example of such a path
and the spin directions associated with the propagation through each
segment\footnote{Note that along each segment the electron undergoes
  many scatterings for $x = l/l_{SO} \ll 1$.}.
Along this entire path the spin can only point up ($+z$-direction),
but never down.  Now, if initially the electrons were unpolarized, the
net out-of-plane polarization in this case would be cancelled by spins
that are initially pointing along the negative $y$-direction.
However, due to the (in-plane) MEE, which itself is subject to strong
fluctuations, e.g. due to conductance fluctuations, there is a finite
in-plane polarization to begin with. The cancellation is therefore
incomplete.  These considerations make plausible that disorder
configurations exist that give rise to strong out-of-plane spin
polarizations.

We next consider the effect of an in-plane magnetic field.
 For $b = 0$ we see that the terms a) and d) in
Eq.~(\ref{eq:fluctuations}) contribute equally to the variance.
However, for $b \neq 0$
the $1/\bar q^2$-divergence is cut off in the Cooperon and we can approximate 
Eq.~(\ref{eq:fluctuations2}) by making use of
\begin{align}
  \label{eq:mag-field-supp}
  s^{ij}_{b} \approx s^{ij} _{b=0} \frac{1}{2} \left[1 +
    1 \Big/ \left( 1 + \frac{32 ( b \tau)^2}{(\bar q^2 + 2 \phi) x^2 }
      \right)
  \right] \, ,
\end{align}
where the first term in Eq.~(\ref{eq:mag-field-supp}) results from the
Diffuson contribution (the term containing $V_{La}, V_{Ra}$ in
Eq.~(\ref{eq:fluctuations})) which is not affected by (moderate) magnetic
fields.

\begin{figure}[t]
  \includegraphics[width = 0.45  \textwidth]{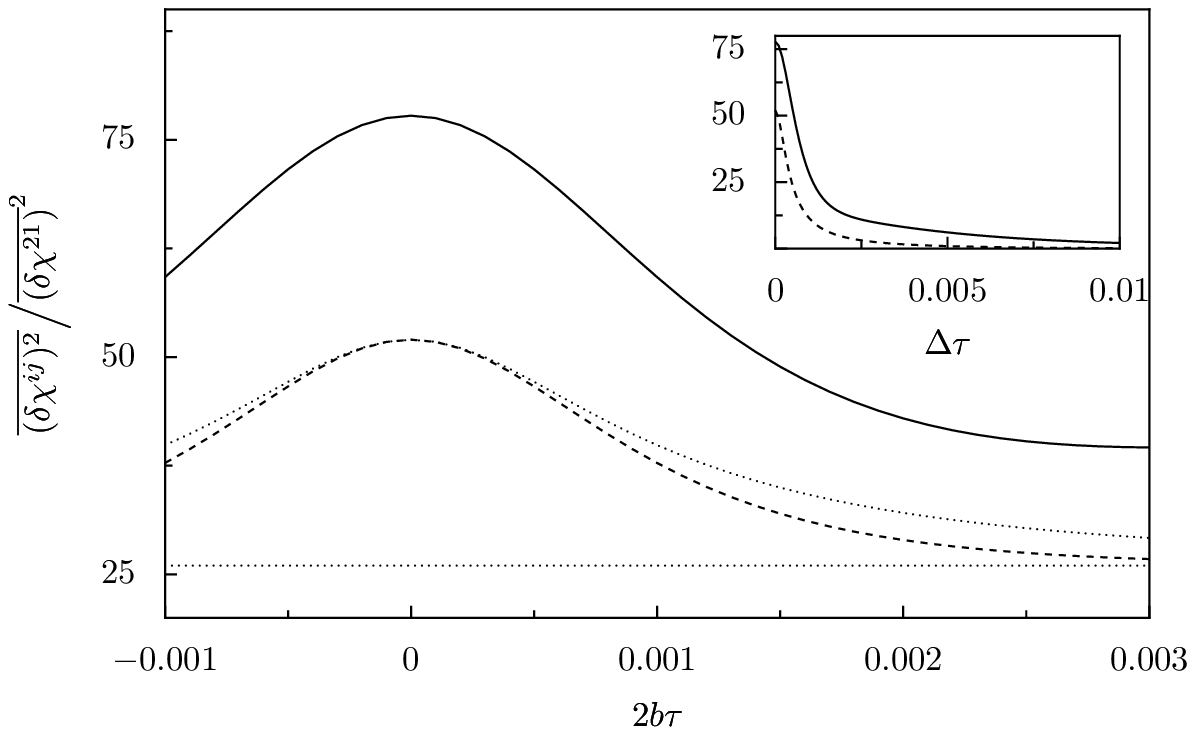}
  \caption{The relative variance of the out-of-plane spin susceptibility
    $\overline{(\delta \chi^{31})^2} / \overline{\delta \chi^{21}}^2$
    is shown as a function of the in-plane magnetic field $2 b \tau$
    for $x=0.1$, $L_\phi = 100 l$, and sample sizes $L = 100 \pi l$
    (solid line) and $L = 200 \pi l$ (dashed line), {\it resp.}  The
    low field approximation, Eq.~(\ref{eq:mag-field-supp}),
    (curved dotted line) and half of the $b=0$-value (straight
    dotted line) are shown for $L = \pi l 100$. Inset:~Variance as a
    function of  $\Delta $ for $x=0.1$,
    $L_\phi = 100 l$, and $L = 100 \pi l$ (solid line)
    and $L = 200 \pi l$ (dashed line).  The suppression of
    $\overline{\delta \chi^{31}}$
    with increasing $b$ and $\Delta $ is described by
    Eqs.~(\ref{eq:mag-field-supp}) and (\ref{eq:gate-volt-supp}).}
\label{fig:variance-gate}
\end{figure}
Unlike the magnetic field, a difference in energies $\Delta $ (e.g. induced by gate voltages) 
leads to a suppression of all terms contributing to $\overline{\delta
  \chi^2}$. This is described by
\begin{align}
  \label{eq:gate-volt-supp}
  s^{ij}_{\Delta } \approx s^{ij}_{\Delta =0}  \Big/ \left( 1 + \frac{4 (
      \Delta  \tau)^2}{(\bar q^2 + 2 \phi)^2 }\right)\, ,
\end{align}
which gives rise to a correlation scale for
susceptibilities at different gate voltages. Indeed, according to
Eq.~(\ref{eq:gate-volt-supp}), we can regard 
$\chi(E_F)$ and $\chi(E_F + \Delta )$ as uncorrelated for $\Delta 
\geq \mathrm{max}\{\phi/\tau, (\pi l/L)^2/\tau\}$.

In conclusion, we find strong mesoscopic fluctuations of the
spin-electric susceptibility in a disordered 2DEG due to Rashba SOI,
giving rise to a large out-of-plane polarization. The predicted values
and dependences on the SOI strength, B-field, dephasing rate, and
Fermi energy are well within experimental reach.  Such spin-dependent
coherence effects, besides being of fundamental interest, might prove
useful in spintronics applications aiming at the electrical control of
spin polarization.

We thank O. Chalaev, O. Tsyplyatyev, and B. Altshuler for 
discussions. We acknowledge financial support from the Swiss NF and
the NCCR Nanoscience Basel.


\end{document}